\documentclass[twocolumns]{aa}

\usepackage{xcolor}
\usepackage[varg]{txfonts}
\usepackage{amsmath}
\usepackage{lscape}
\usepackage{multirow}
\usepackage{pdflscape}
\usepackage{hyperref} 
\usepackage{adjustbox}

\def\kms{\hbox{km$\;$s$^{-1}$}}
\def\cm3{\hbox{cm$^{-3}$}}
\def\deg{\hbox{$^{\circ}$}}

\begin{document}

\title{Direct evidence for magnetic reconnection at the boundaries of magnetic switchbacks with Parker Solar Probe}

\author{C. Froment\inst{1}
\and V. Krasnoselskikh\inst{1,2}
\and T. Dudok de Wit\inst{1}
\and O. Agapitov\inst{2}
\and N. Fargette\inst{3}
\and B. Lavraud\inst{3,4}
\and A. Larosa\inst{1}
\and M. Kretzschmar\inst{1}
\and V. K. Jagarlamudi\inst{1,5}
\and M. Velli\inst{6}
\and D. Malaspina\inst{7,8}
\and P. L. Whittlesey\inst{2}
\and S. D. Bale\inst{2,9,10,11}
\and A. W. Case\inst{12}
\and K. Goetz\inst{13}
\and J. C. Kasper\inst{12,14,15}
\and K. E. Korreck\inst{12}
\and D. E. Larson\inst{2}
\and R. J. MacDowall\inst{16}
\and F. S. Mozer\inst{9}
\and M. Pulupa\inst{2}
\and C. Revillet\inst{1}
\and M. L. Stevens\inst{12}}

\institute{LPC2E, CNRS/University of Orl\'eans/CNES, 3A avenue de la Recherche Scientifique, Orl\'eans, France\\
\email{clara.froment@cnrs-orleans.fr}
\and
Space Sciences Laboratory, University of California, Berkeley, CA 94720-7450, USA
\and
Institut de Recherche en Astrophysique et Planétologie, CNRS, UPS, CNES, Toulouse, France
\and
Laboratoire d'Astrophysique de Bordeaux, Univ. Bordeaux, CNRS, B18N, allée Geoffroy Saint-Hilaire, 33615 Pessac, France
\and
National Institute for Astrophysics-Institute for Space Astrophysics and Planetology, Via del Fosso del Cavaliere 100, I-00133 Roma, Italy
\and
Department of Earth, Planetary, and Space Sciences, UCLA, Los Angeles, CA, 90095, USA
\and
Laboratory for Atmospheric and Space Physics, University of Colorado, Boulder, CO 80303, USA
\and
Astrophysical and Planetary Sciences Department, University of Colorado, Boulder, CO 80303, USA
\and
Physics Department, University of California, Berkeley, CA 94720-7300, USA
\and
The Blackett Laboratory, Imperial College London, London, SW7 2AZ, UK
\and
School of Physics and Astronomy, Queen Mary University of London, London E1 4NS, UK
\and
Smithsonian Astrophysical Observatory, Cambridge, MA, 02138, USA
\and
School of Physics and Astronomy, University of Minnesota, Minneapolis, MN 55455, USA
\and
Climate and Space Sciences and Engineering, University of Michigan, Ann Arbor, MI 48109, USA
\and
BWX Technologies, Inc., Washington, DC 20002, USA
\and
Solar System Exploration Division, NASA/Goddard Space Flight Center, Greenbelt, MD, 20771, USA
}

\date{Received October, 30 2020 / Accepted January 15, 2021}

\abstract
{The first encounters of Parker Solar Probe (PSP) with the Sun revealed the presence of ubiquitous localised magnetic deflections in the inner heliosphere; these structures, often called switchbacks, are particularly striking in solar wind streams originating from coronal holes.}
{We report the direct piece of evidence for magnetic reconnection occurring at the boundaries of three switchbacks crossed by PSP at a distance of 45 to 48 solar radii to the Sun during its first encounter.}
{We analyse  the magnetic field and plasma parameters from the FIELDS and Solar Wind Electrons Alphas and Protons (SWEAP) instruments.}
{The three structures analysed all show typical signatures of magnetic reconnection.  The ion velocity and magnetic field are first correlated and then  anti-correlated at the inbound and outbound edges of the bifurcated current sheets with a central ion flow jet. Most of the reconnection events have a strong guide field and moderate magnetic shear, but one current sheet shows indications of quasi anti-parallel reconnection in conjunction with a magnetic field magnitude decrease by $90\%$.} 
{Given the wealth of intense current sheets observed by PSP, reconnection at switchback boundaries appears to be rare. However, as the switchback boundaries accomodate currents, one can conjecture that the geometry of these boundaries offers favourable conditions for magnetic reconnection to occur. Such a mechanism would thus contribute in reconfiguring the magnetic field of the switchbacks, affecting the dynamics of the solar wind and eventually contributing to the blending of the structures with the regular wind as they propagate away from the Sun.}


\titlerunning{Magnetic reconnection at a switchback boundary}
\authorrunning{Froment et al.}
\maketitle

\section{Introduction}\label{sec:intro}
The solar atmosphere constantly releases a stream of particles, known as the solar wind. How these particles are accelerated to the interplanetary medium is still an open question. Parker Solar Probe \citep[PSP,][]{fox15} is the first spacecraft to go close enough to the Sun to sample the in situ characteristics of the pristine solar wind in order to tackle this question. During the first encounter of PSP with the Sun, in November 2018, the spacecraft was embedded in a slow-wind stream originating from a small coronal hole \citep{riley_predicting_2019, Badman2020, kim_predicting_2020, szabo_heliospheric_2020} and made an approach at 35.7 solar radii.
These observations revealed the ubiquitous presence of deflections in the magnetic field of the solar wind \citep{bale19}, which are particularly pronounced in the variation of its radial component. Although most deflections (from the prevalent sunward polarity) are of a few tens of degrees, some correspond to full reversals \citep{ddw20, mozer_switchbacks_2020}. Their duration spans from a few seconds to a few hours \citep{ddw20}. For these structures the electron pitch-angle distributions show that the strahl, the high energy electron population emanating from the Sun, follows the orientation of the magnetic field \citep{kasper19}. This suggest that these \textquoteleft
switchbacks\textquoteright~are localised twists in the magnetic field and not simple polarity reversals or closed loops. Cross helicity measurements, in addition, show that magneto-hydrodynamic waves follow the magnetic field inside the switchbacks \citep{mcmanus_cross_2020}. 
This magnetic field twist is coupled with a proton velocity enhancement that is most of the time, but not always, of the order of the local Alfv\'en velocity. As shown by \citet{larosa_switchbacks_2020}, velocity enhancements inside switchbacks can be only of a few percent of the velocity outside of the structures. The statistics of these enhancements show a continuum of fractions of the local Alfv\'en velocity. 

Structures similar to switchbacks were observed before in the solar wind \citep[e.g][]{balogh_heliospheric_1999, horbury_short_2018}. However, the new observations from PSP revealed persistent features, populating a significant fraction of the measurements. Such short events had also never been observed before. The scattered observation of switchbacks further away from the Sun could be due to their deterioration as they travel, even though these structures seem to be able to live long enough in the slow solar wind to be generated close to the Sun and observed at a few tens of solar radii \citep{tenerani_magnetic_2020a}. The ubiquity of switchbacks in PSP observations (also during subsequent perihelia) could also be due to the co-rotation of the probe with the Sun during the encounters, allowing the probe to meet a larger number of them, even if they were emitted by a rather localised source on the Sun  \citep{horbury_sharp_2020}. We also note that there is presently no consensus between the few statistical studies that have quantified the degree of occurrence of switchbacks with the radial distance from the Sun \citep[][]{tenerani_magnetic_2020b, velli_agu_2020}.
 
Observational studies invoke the formation of these structures either by processes occurring deep in the solar atmosphere \citep{ddw20, krasnoselskikh_localized_2020, macneil_parker_2020, woodham_enhanced_2020} or directly in the solar wind \citep{ruffolo_shear-driven_2020}.
These two types of origins have also been explored on the numerical and theoretical side: on the one hand, via interchange reconnection in the corona \citep{fisk_global_2020,drake_are_2020} or jets in the solar atmosphere \citep{roberts_simulated_2018, sterling_coronal-jet-producing_2020, he_possible_2020}; on the other hand, via waves growth generated by the expansion of the structure away from the Sun \citep{squire_-situ_2020} or shear-driven turbulence in the solar wind \citep{ruffolo_shear-driven_2020}. 
As of today, the mechanisms leading to the formation of  switchbacks is still unclear and it is unknown whether different populations of switchbacks might be generated by distinct mechanisms. Their omnipresence, however, shows that they could play an important role in the dynamics and heating of the solar wind.

Several interesting properties of the switchbacks arise from the study of their boundaries. The rotation of the magnetic field at the boundary of switchbacks make them current sheets by definition. Via a superposed epoch analysis, \citet{farrell_magnetic_2020} highlight that events with a rapid deflection in the magnetic field show a clear decrease in the magnetic field magnitude at their boundaries (7\% - 8\%). The authors conjecture that these dropouts are caused by a diamagnetic current created at the boundaries by a pressure gradient in the region where the magnetic field is rotating. This kind of magnetic field dropout coupled with a proton density increase at switchback boundaries were clearly present in several studies \citep[e.g][]{bale19,krasnoselskikh_localized_2020, agapitov20}. \citet{krasnoselskikh_localized_2020} have also reported strong currents at switchback boundaries, which is confirmed by the statistical analysis conducted by \citet{larosa_switchbacks_2020}.

Current sheets are privileged regions for magnetic reconnection to occur. In the case of magnetic reconnection, the magnetic field topology is reconfigured to a simpler configuration, converting magnetic energy into thermal and kinetic energies \citep[e.g][]{cassak16,zweibel09,zweibel_perspectives_2016}. Particles are ejected away from the reconnection site at a velocity close to the Alfv\'en velocity, and the plasma is locally heated. Reconnection in the solar wind is typically detected through the observation of an ion flow jet within a bifurcated current sheet \citep{gosling_direct_2005}. The probability of crossing the reconnection site, the so-called diffusion region, is low due to its small spatial scale (of the order of the ion inertial length $\mathrm{d_i}$). A correlation between the evolution of the magnetic field and the particle density, later followed by an anti-correlation, or the other way around, is typically observed when reconnection exhausts are crossed. Such behaviour is consistent with Alfv\'en waves propagating along the magnetic field, away from the reconnection site on either side of the exhaust, as demonstrated by \citet{gosling_direct_2005}. In PSP observations, reconnection exhausts have been reported at the heliospheric current sheets (HCS), in interplanetary coronal mass ejections (ICME), and in some isolated current sheets, but not so far at the most frequent current sheets observed which are those that bound switchbacks \citep[e.g.][]{phan_parker_2020,lavraud_heliospheric_2020, fargette_magnetic_2020}.

In this paper, we analyse the boundaries of three switchbacks that were crossed during the first solar encounter of PSP. At least five out of six current sheets show direct evidence of magnetic reconnection. They were observed on November 1 and November 2, 2018, that is to say five and four days before perihelion. At that time, PSP was located between 48 and 45 solar radii from the Sun. The observations are presented in Sect.~\ref{sec:observations}. After the detailed analysis of the three events in Sect.~\ref{sec:analysis}, we summarise their properties and discuss the implications of our observations on the switchbacks dynamics and propagation in the solar wind in Sect.~\ref{sec:conclusion}.

\begin{figure*}
	\resizebox{\hsize}{!}
	{\begin{tabular}{cc} 
		\multicolumn{2}{c}{\includegraphics[width=2.5\textwidth]{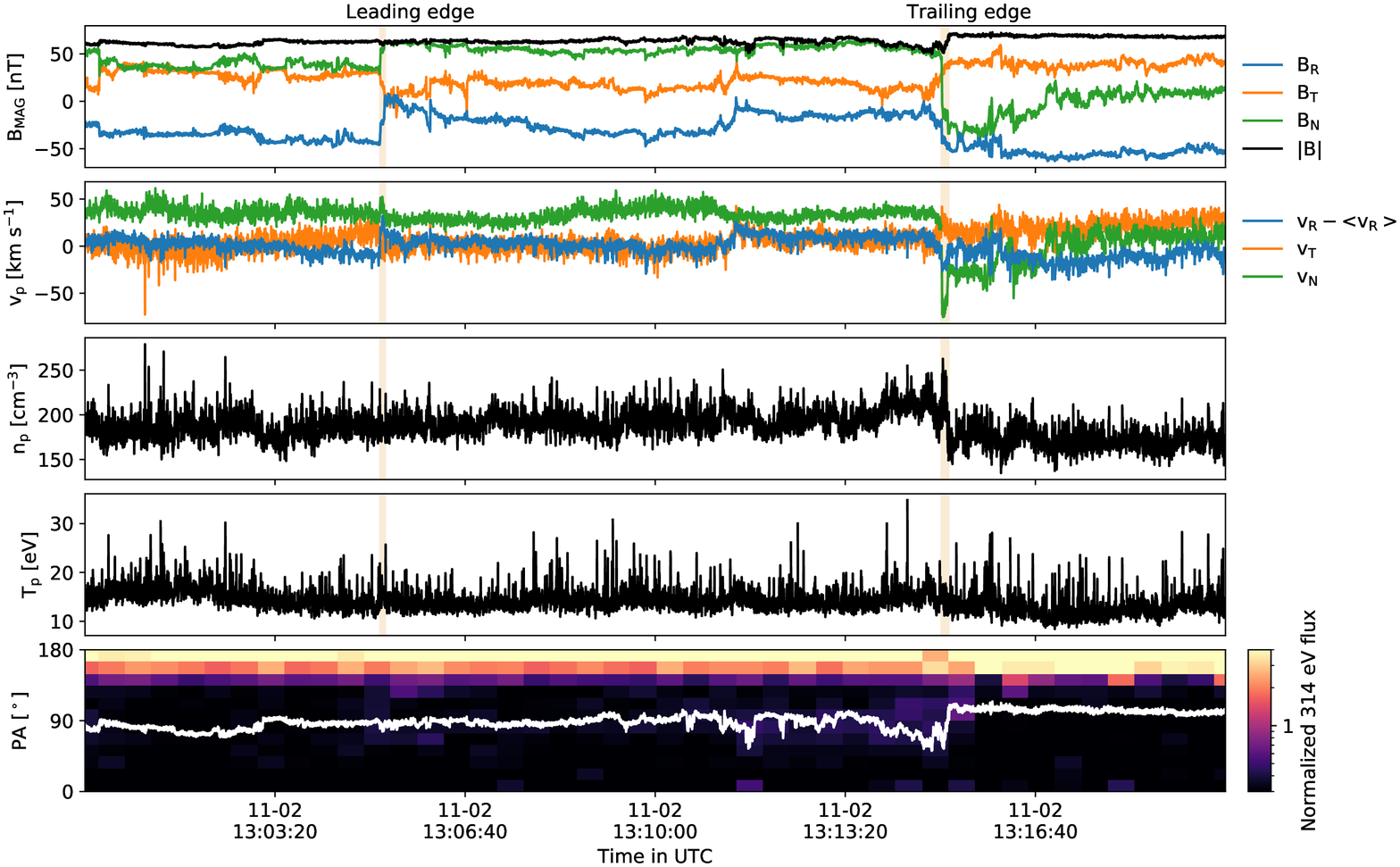}} \\
		\includegraphics{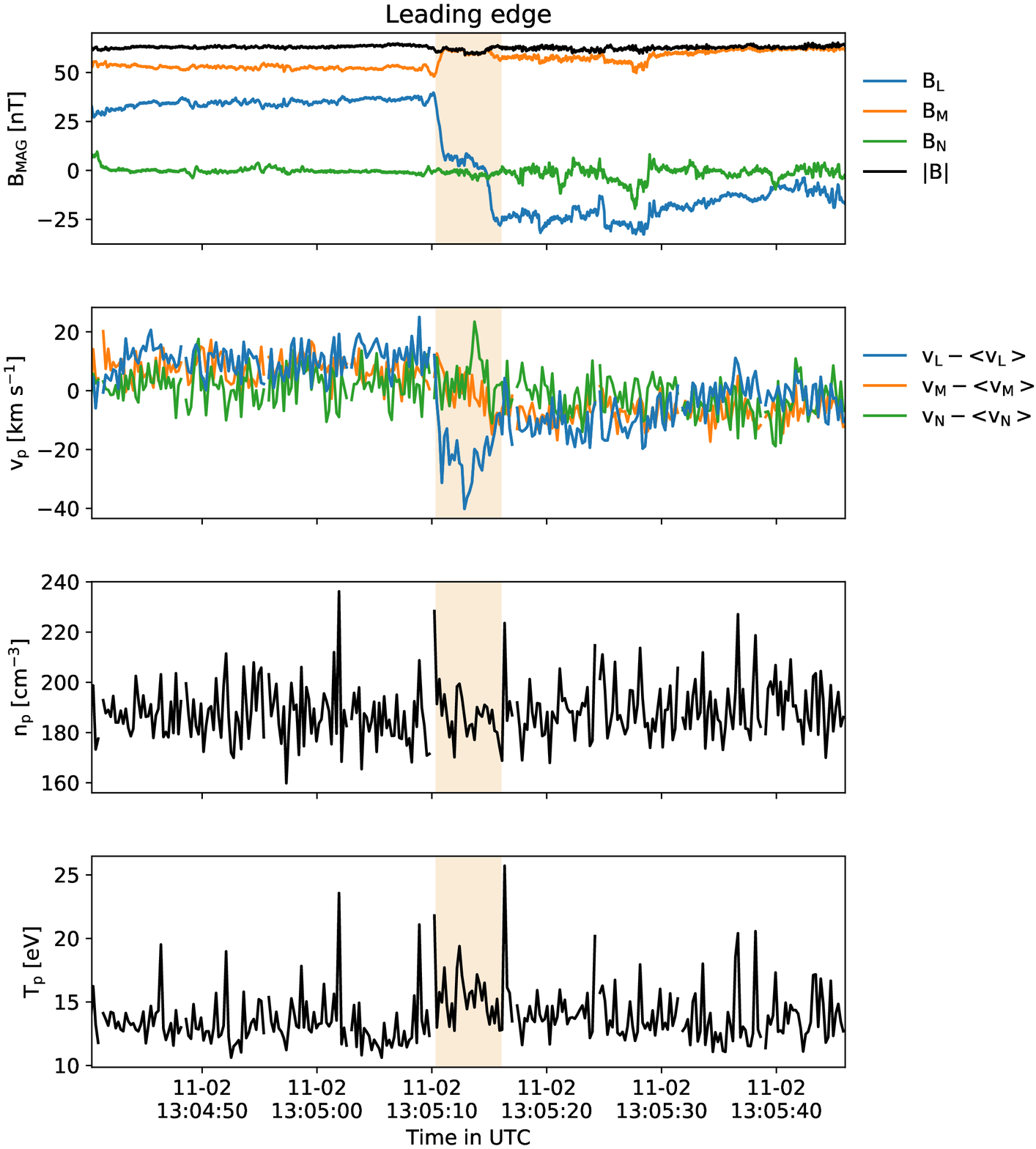} & \includegraphics{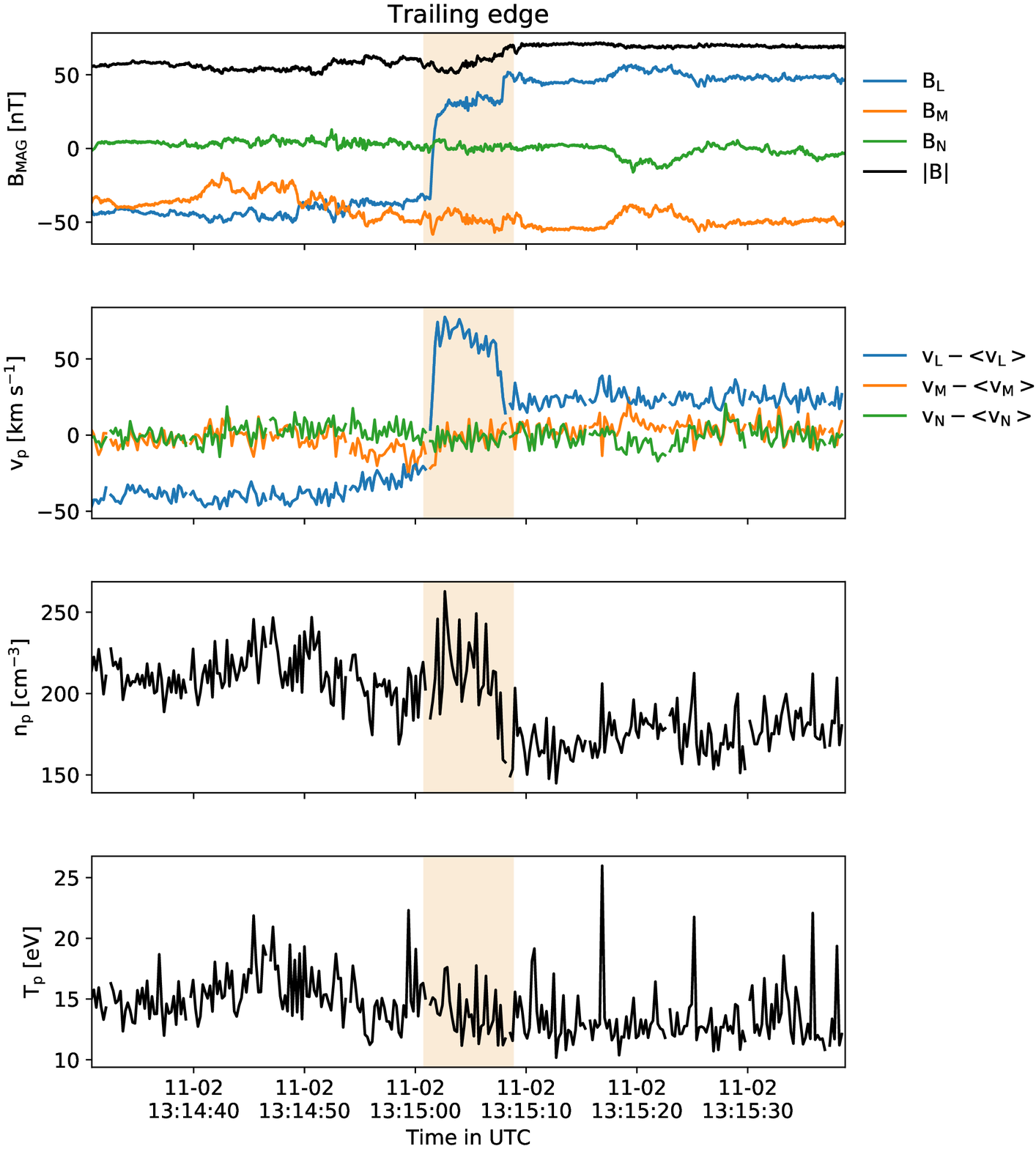}
	\end{tabular}
	}
	\caption{Event~1 on November 2, 2018, starting at 13:05:09 UT. The shaded areas delimit the boundaries of the switchback, noted as leading and trailing edges. First panel: DC magnetic field measurements from FIELDS/MAG in the RTN frame. Second panel: Proton velocity from SWEAP/SPC, also in the RTN frame. The average value of the radial velocity ($<\mathrm{v_R}> = $ 320 \kms) was removed for easier visualisation. 
	Third panel: Proton density from SWEAP/SPC. Fourth panel: Radial proton temperature from SWEAP/SPC. Fifth panel: Pitch angle distribution of 314 eV electrons from SWEAP/SPAN-E. The combined flux (two sensors) was normalised by the averaged flux. The white curve represents the magnitude of the magnetic field in arbitrary units. Bottom panels: The two sets of plots at the bottom are excerpts of the leading and trailing edges of the switchback; they represent the same variables (except for the pitch-angle distribution) in the same order. For these two sets of plots, the magnetic field and proton velocity are represented in the local current sheet coordinates LMN. The average value of the velocity for each component was removed for easier visualisation.}
	\label{fig:context_event_1}
\end{figure*}

\begin{figure*}
	\resizebox{\hsize}{!}
	{\begin{tabular}{cc} 
        \includegraphics{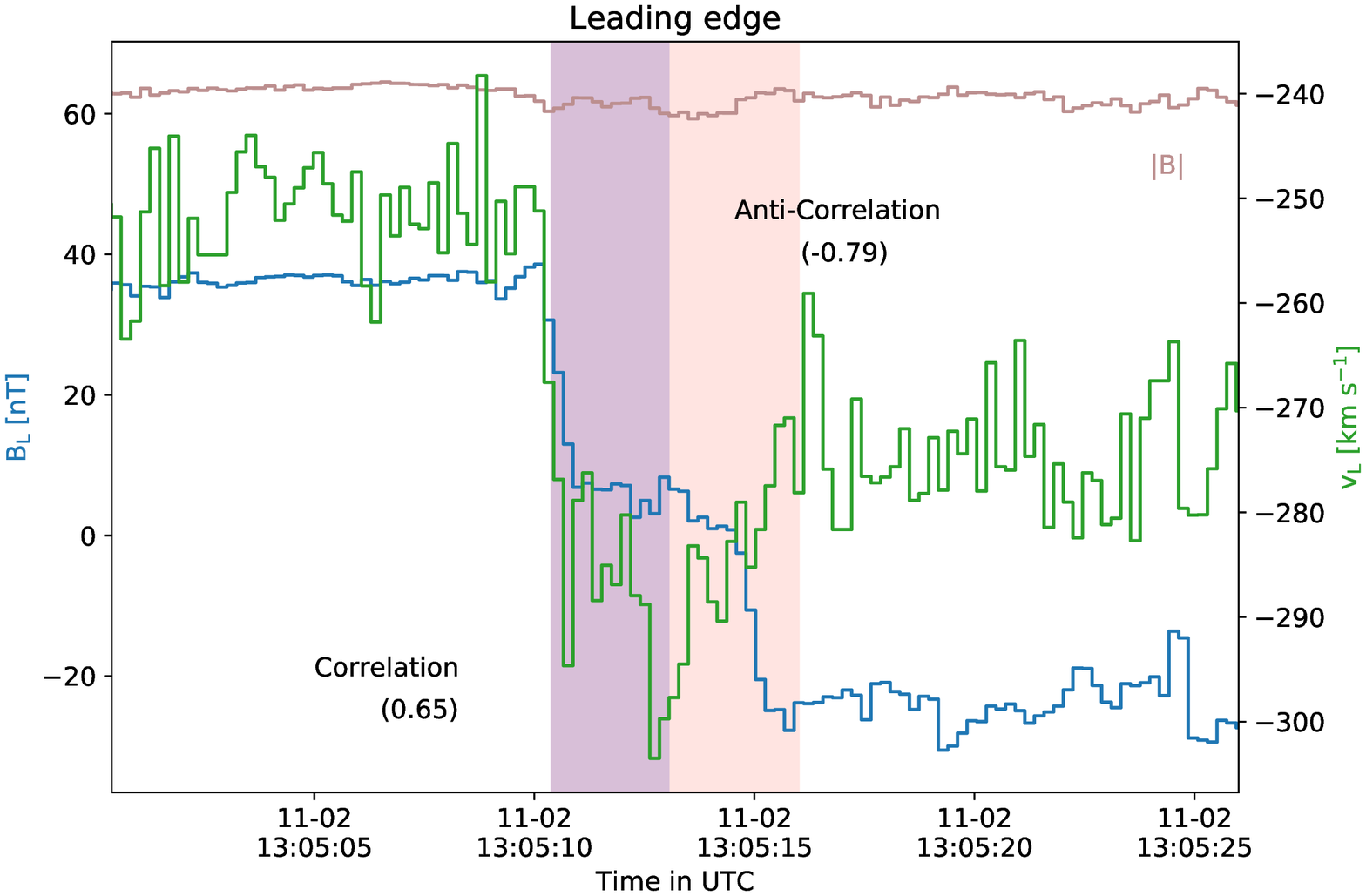} & \includegraphics{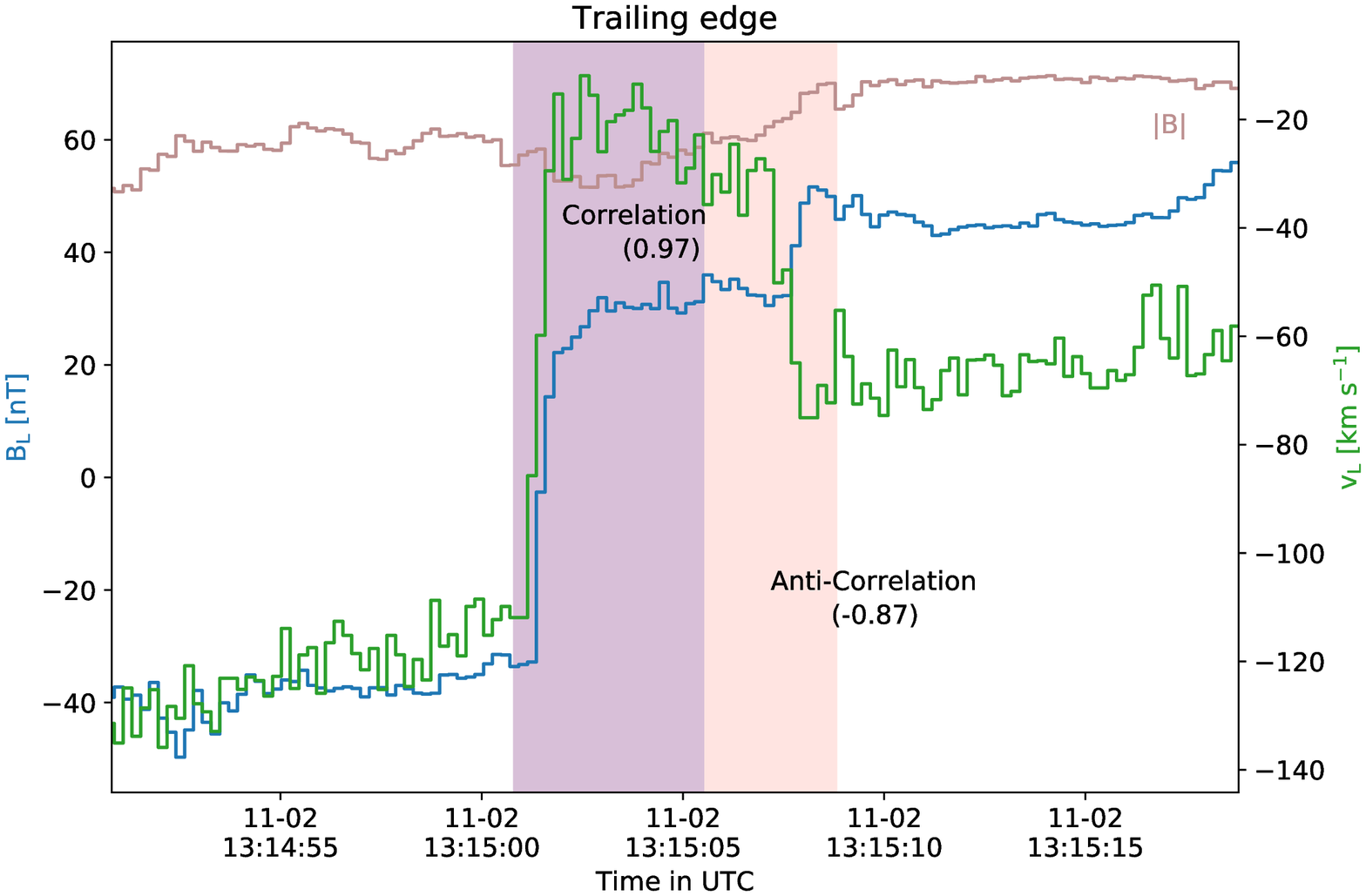}
	\end{tabular}
	}
	\caption{Correlation and anti-correlation of $\mathrm{B_L}$ and $\mathrm{v_L}$ at the boundaries of event 1. The leading edge is represented in the left panel and trailing edge in the right panel. Evolution of $\mathrm{B_L}$ and $\mathrm{v_L}$, component along the main direction of the current sheets. The Pearson correlation coefficients shown were computed in the coloured time windows, chosen to maximise the correlation values. The light brown line is the magnitude of the magnetic field.}
	\label{fig:correlation_boundaries_event_1}
\end{figure*}

\begin{figure*}
	\resizebox{\hsize}{!}
	{\begin{tabular}{cc} 
		\multicolumn{2}{c}{\includegraphics[width=2.5\textwidth]{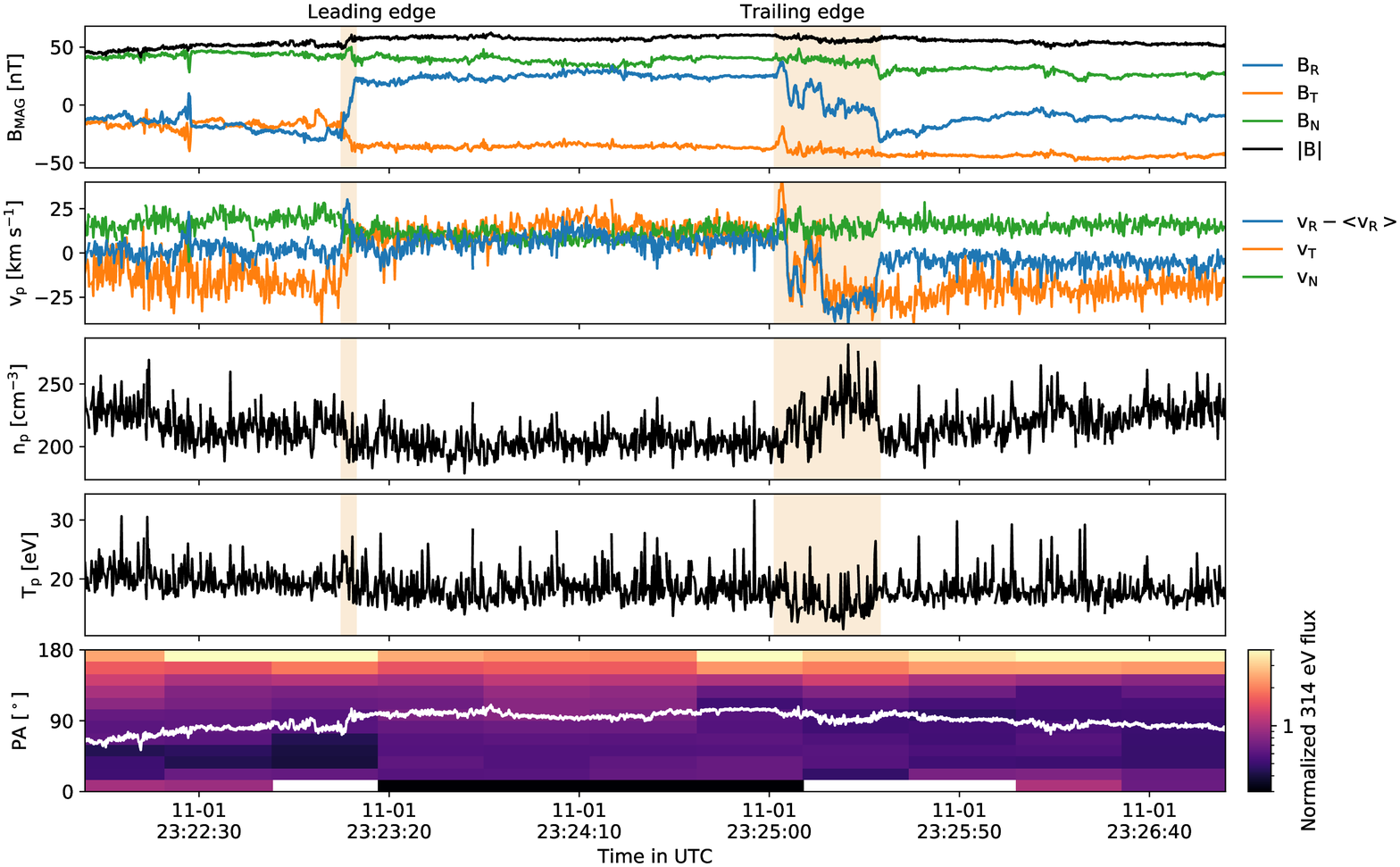}} \\
		\includegraphics{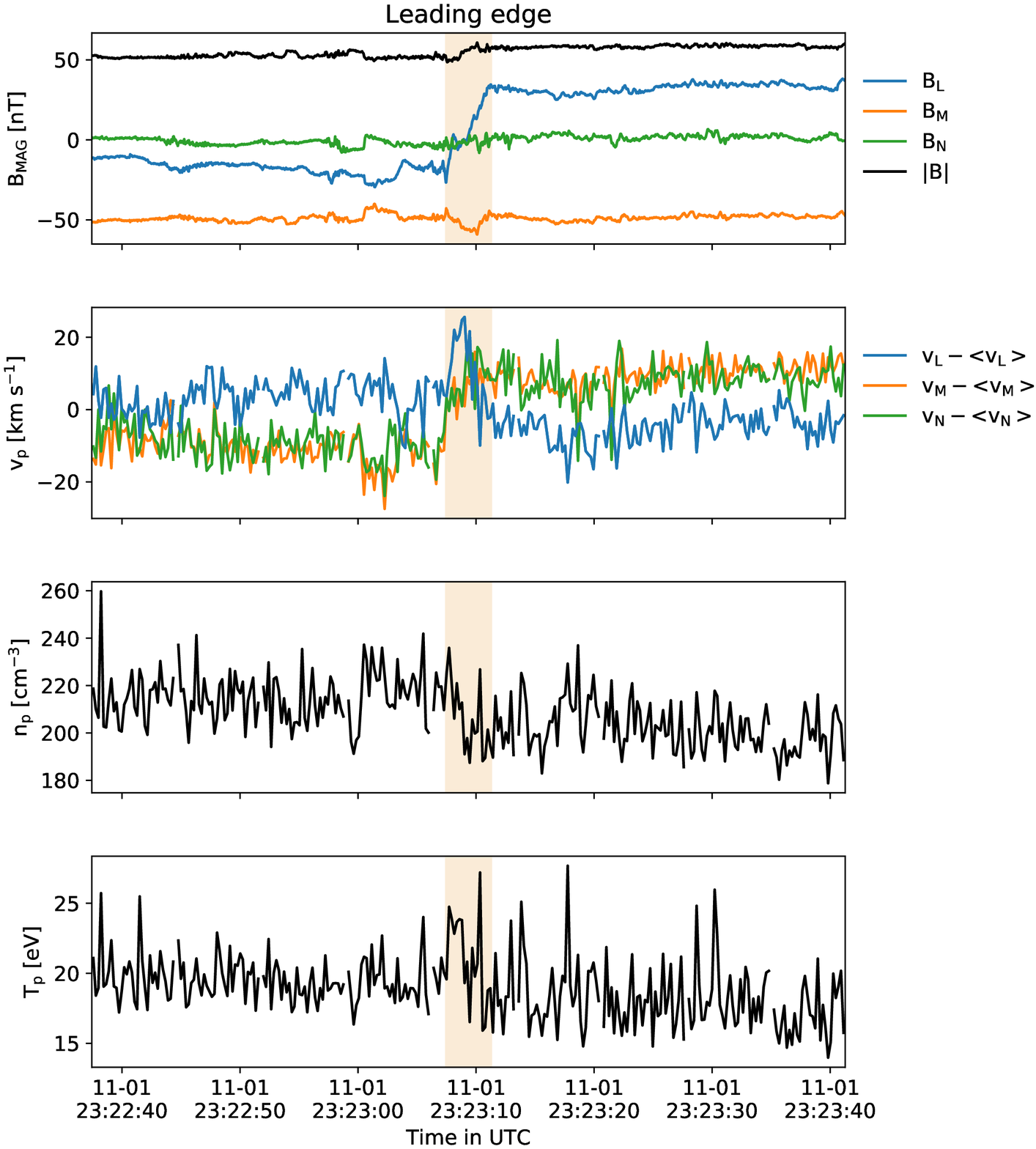} & \includegraphics{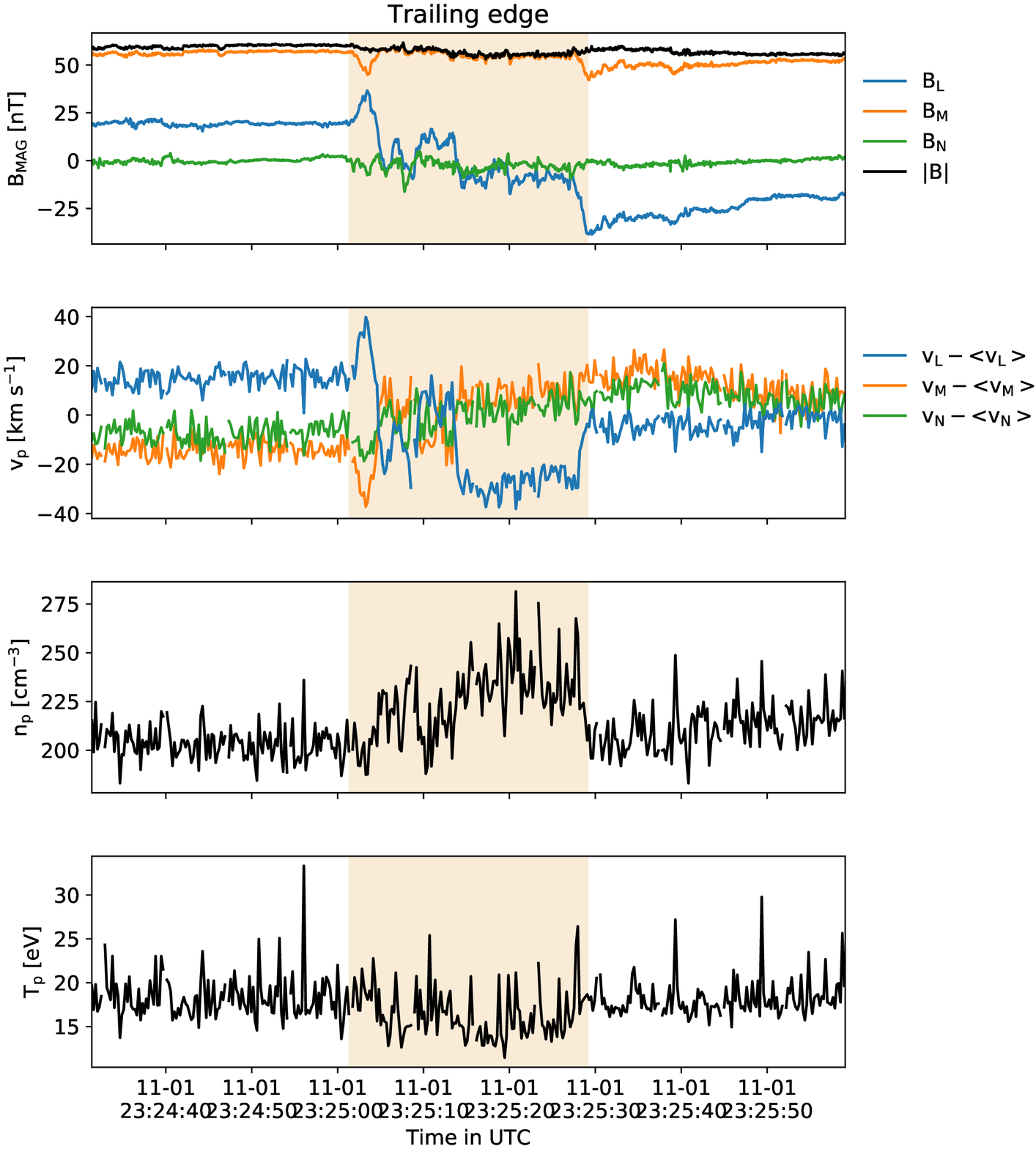}
	\end{tabular}
	}
	\caption{Same as Fig.~\ref{fig:context_event_1}, but for event 2 on November 1, 2018, starting at 23:23:04 UT. The average value of the radial velocity ($<\mathrm{v_R}> = $ 342 \kms) was removed for easier visualisation.}
	\label{fig:context_event_2}
\end{figure*}

\begin{figure}
    \centering
	\includegraphics[width=\linewidth]{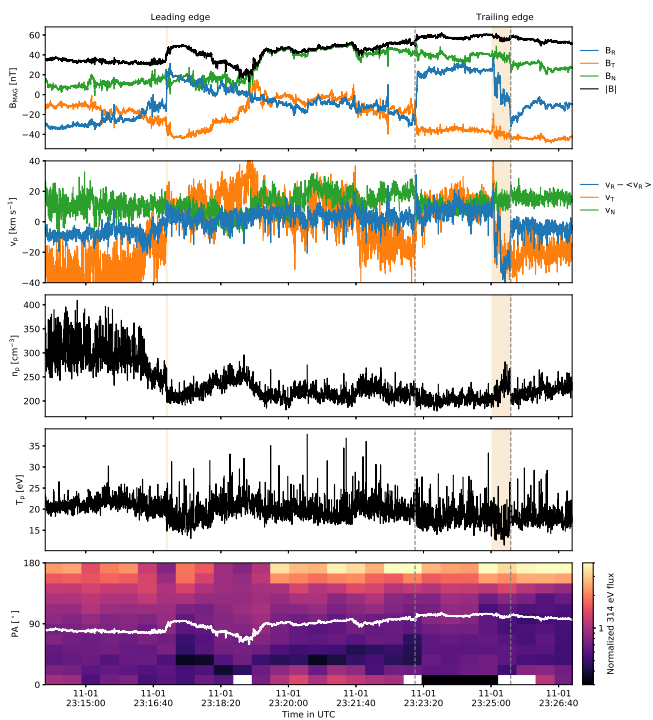} 
	\caption{Larger switchback in which event 2 is embedded. The four panels show the same variables as in the first figure of Fig.~\ref{fig:context_event_2}. The sub-structure noted as event 2 is delimited here by two vertical dotted grey bars.}
	\label{fig:context_event_2_large}
\end{figure}

\begin{figure*}
	\resizebox{\hsize}{!}
	{\begin{tabular}{cc} 
        \includegraphics{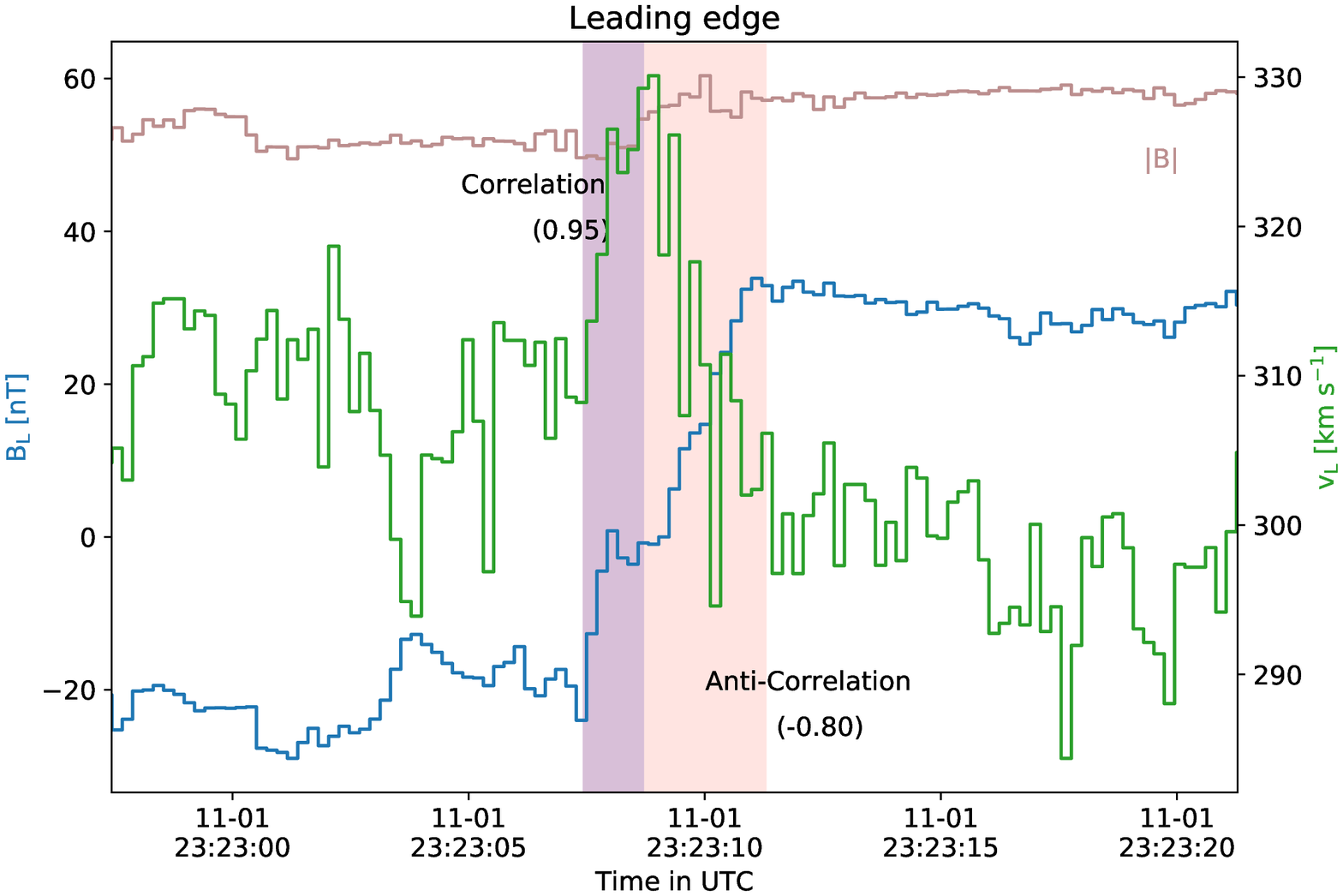} & \includegraphics{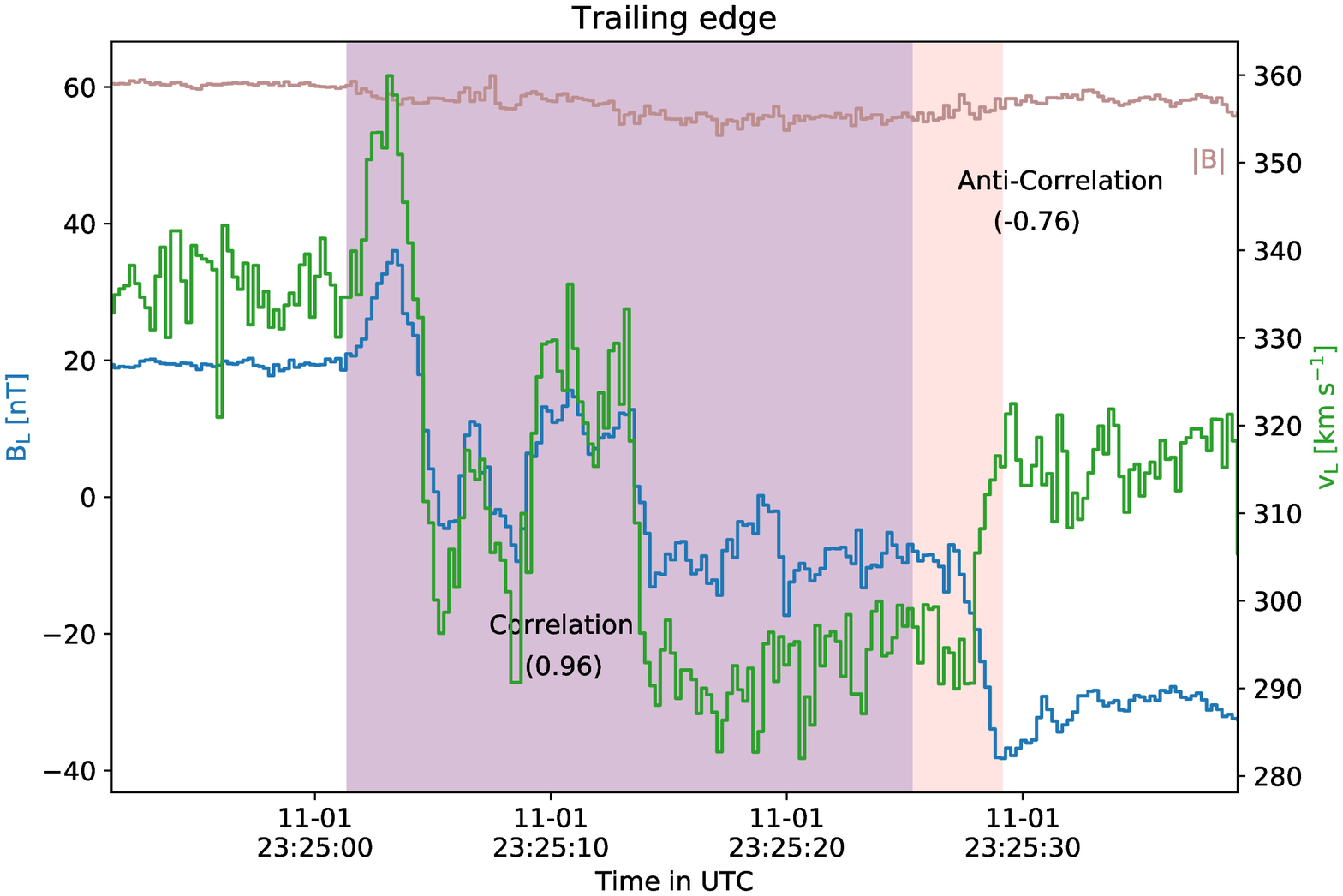}
	\end{tabular}
	}
	\caption{Same as Fig.~\ref{fig:correlation_boundaries_event_1} but for event 2. Correlation and anti-correlation of $\mathrm{B_L}$ and $\mathrm{v_L}$ at both boundaries.}
	\label{fig:correlation_boundaries_event_2}
\end{figure*}

\begin{figure*}
	\resizebox{\hsize}{!}
	{\begin{tabular}{cc} 
		\multicolumn{2}{c}{\includegraphics[width=2.5\textwidth]{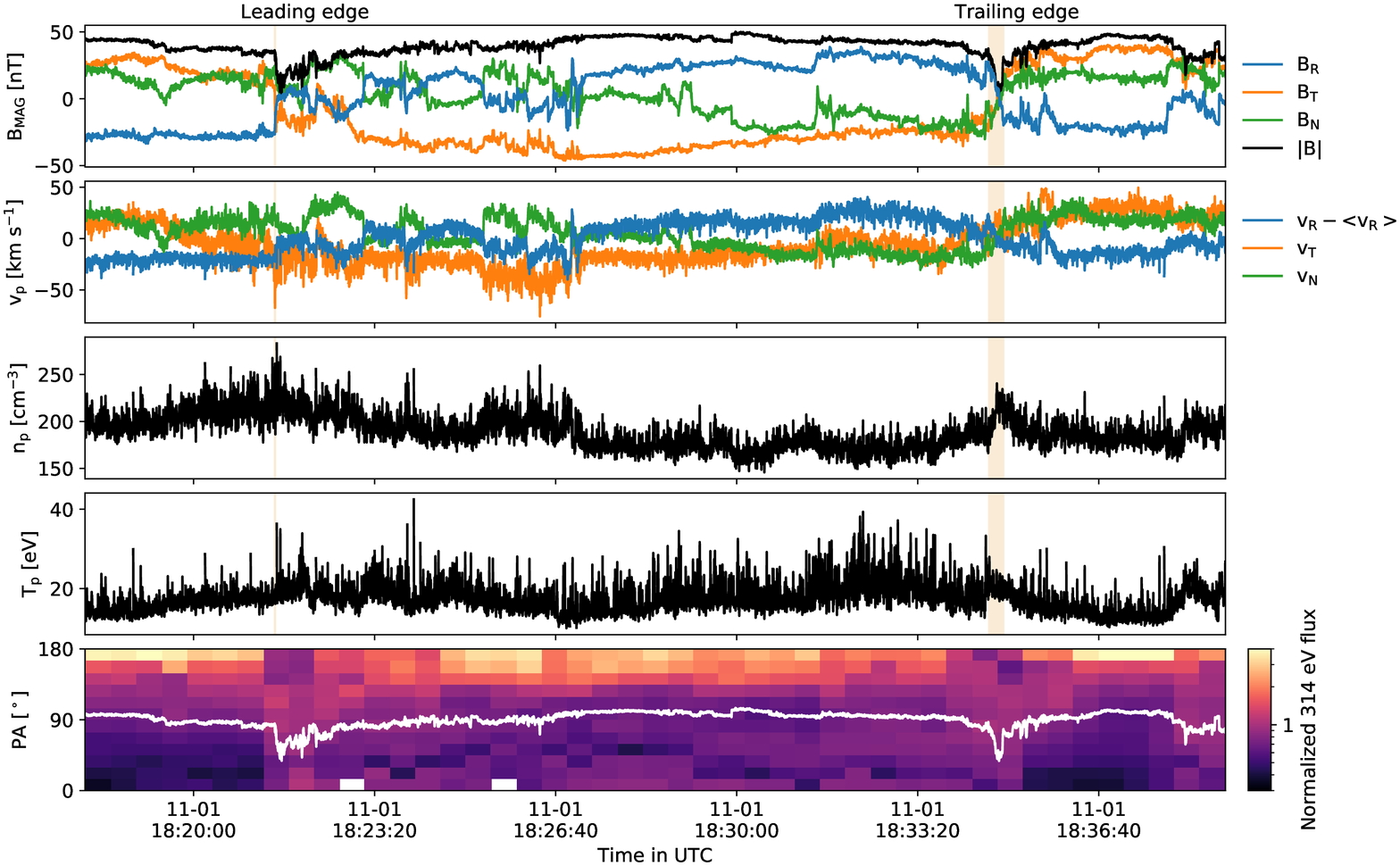}} \\
		\includegraphics{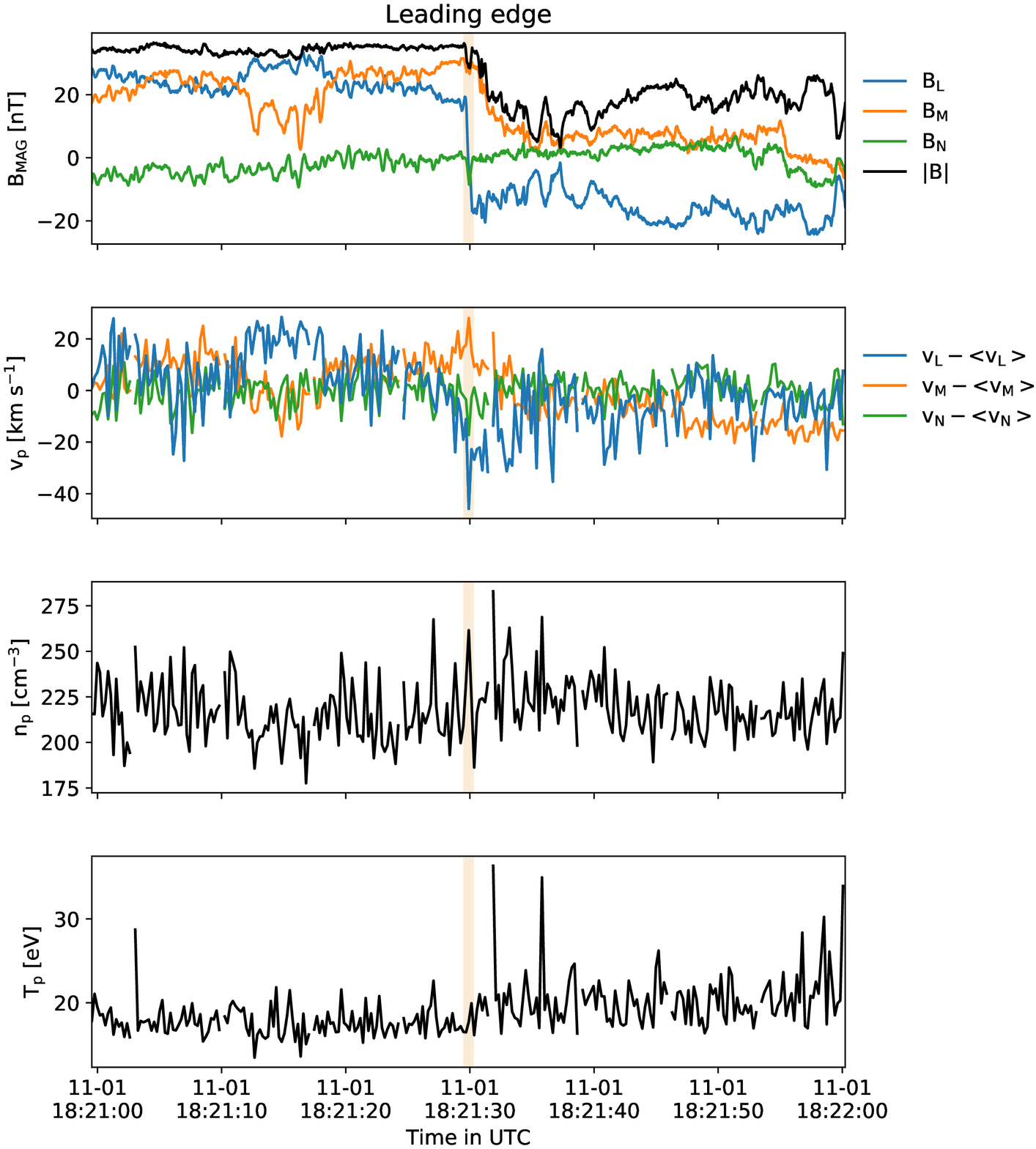} & \includegraphics{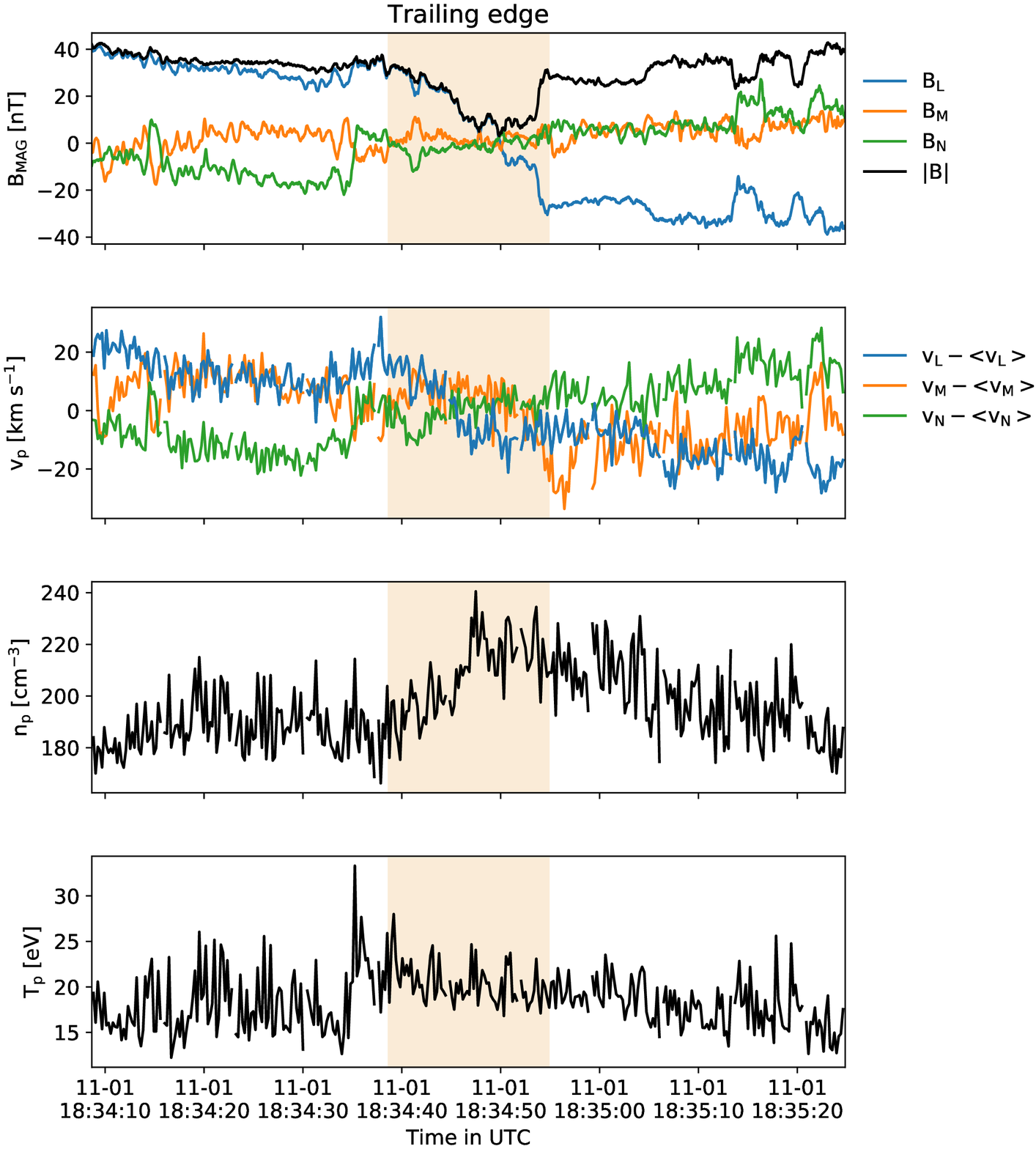}
	\end{tabular}
	}
	\caption{Same as Fig.~\ref{fig:context_event_1} but for event 3 on November 1, 2018, starting at 18:21:28 UT. The average value of the radial velocity ($<\mathrm{v_R}> = $ 341 \kms) was removed for easier visualisation.}
	\label{fig:context_event_3}
\end{figure*}

\begin{figure*}
	\resizebox{\hsize}{!}
	{\begin{tabular}{cc} 
        \includegraphics{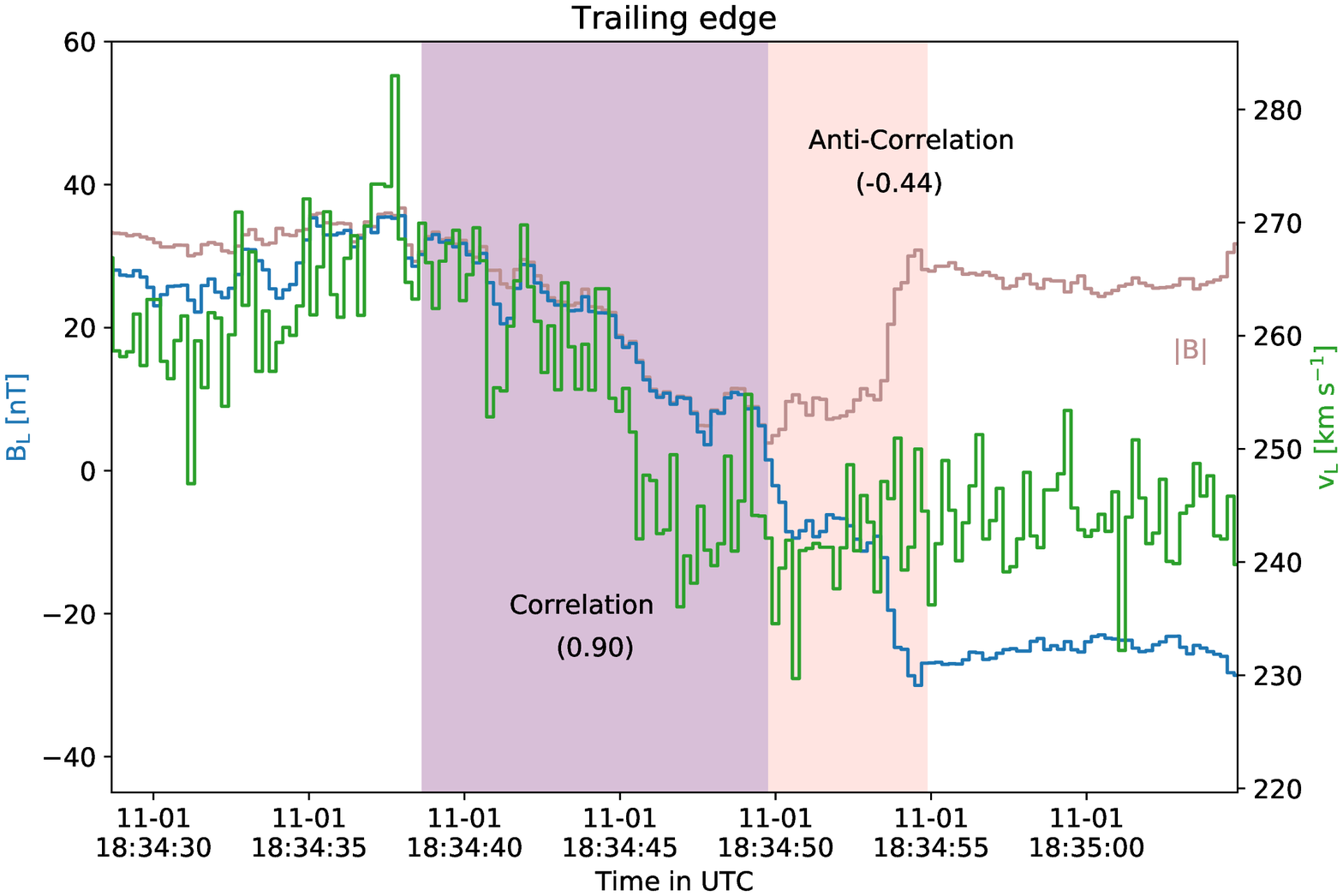} & \includegraphics{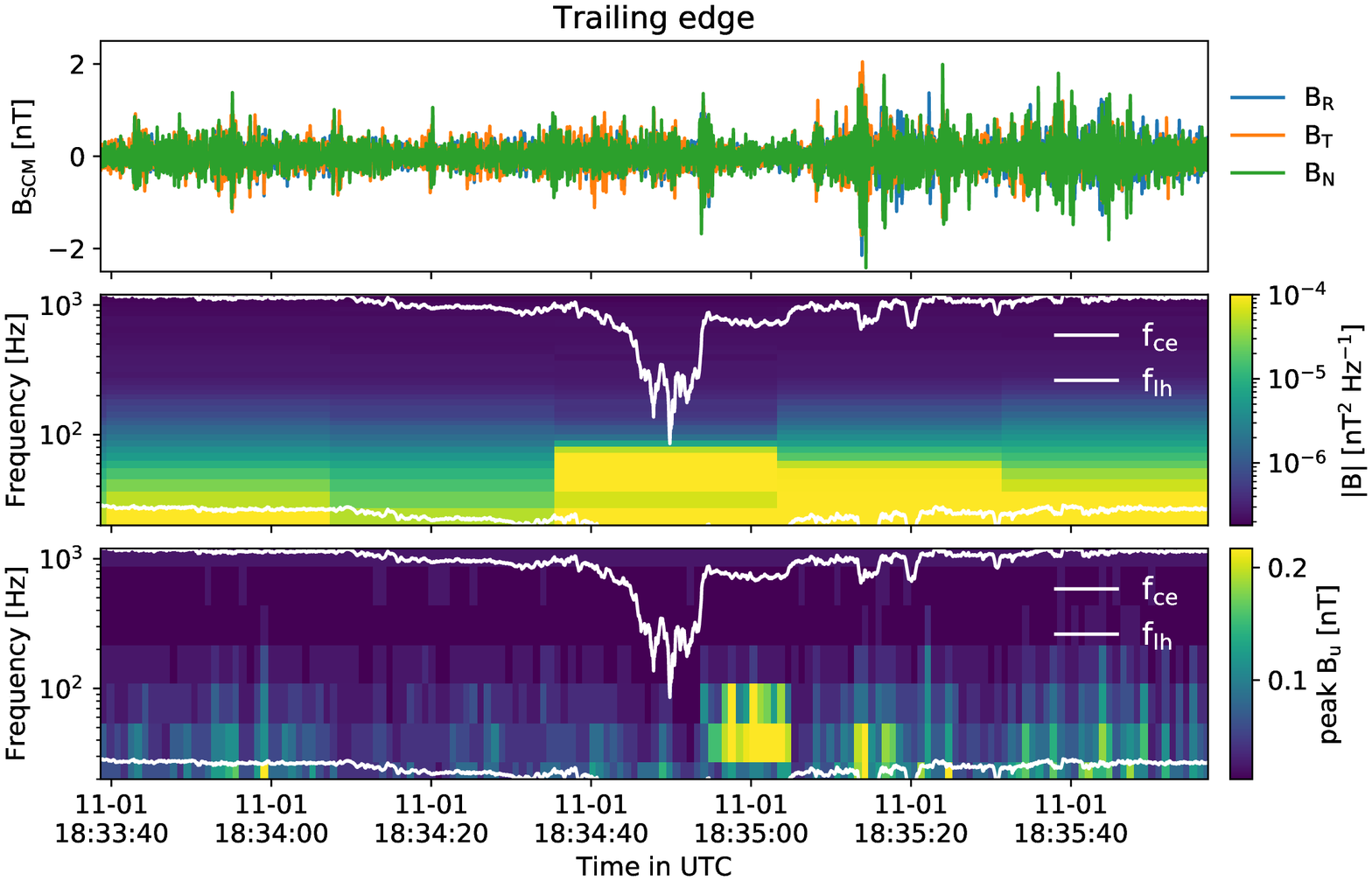}
	\end{tabular}
	}
	\caption{Evidence of reconnection at the trailing edge of event 3. The left panel follows the same representation as in Fig.~\ref{fig:correlation_boundaries_event_1} but for event 3 and shows a correlation and anti-correlation of $\mathrm{B_L}$ and $\mathrm{v_L}$. The right panels show the whistler waves detected right after PSP crossed the current sheet. First panel: SCM waveforms in the RTN coordinates system. Second panel: Cross-spectral measurement (trace of the spectral matrix). Third panel: BPF measurements (one component in the SCM sensor frame). For the last two panels, the white lines represent the lower hybrid and electron cyclotron frequencies, respectively.}
	\label{fig:correlation_boundaries_event_3}
\end{figure*}

%%%%%%%%%%%%%%%%%%%%%%%%%%%%%%%%%

\section{PSP observations}
\label{sec:observations}

\subsection{Data}
We used magnetic field measurements made by the FIELDS suite of instruments \citep{Bale2016}. The vector magnetic field was measured from DC to several tens of hertz by the fluxgate magnetometer (MAG) while magnetic fluctuations above 10~Hz were measured by the Search-Coil Magnetometer \citep[SCM,][]{jannet_2020}.
Our analysis of wave activity (Sect.~\ref{sec:event_3}) and its properties is based on waveforms (73.24 samples per second, 3 components), auto- and cross-spectra (cadence of 28 seconds, full spectral matrix), and band-pass filter (BPF) time series (cadence of 0.87 seconds, one component only: $u$ in the SCM sensor frame) of the search-coil. BPF time series represent the amplitude of the wavefield in specific spectral bands. All these data products are provided by the Digital Fields Board \citep[DFB,][]{malaspina_digital_2016}. The sampling rate of the waveforms corresponds to the survey cadence during the early part of the solar encounter phase. During the close encounter phase, this cadence increases fourfold. However, none of the switchback boundaries observed during the close encounter show as clear evidence for reconnection as the ones that we present below.

The proton velocity, density, and temperature are provided by the Solar Wind Electrons Alphas and Protons (SWEAP) suite \citep{Kasper2016}. SPC Faraday cups \citep[][]{case_solar_2020} provide  moments of the reduced distribution function of ions: density, velocity, and radial component of the thermal velocity. From the latter, we estimated the radial proton temperature: $T_R = \frac{1}{2}\frac{\mathrm{M_i}}{\mathrm{k_B}}\delta \mathrm{v}^2$ [eV], where $\mathrm{k_B}$ is the Boltzmann constant, $\mathrm{M_i}$ is the mass of protons, and $\delta\mathrm{v}^2$ is the proton radial thermal velocity. These data come with different quality flags; we only used  good quality data. Their cadence is 0.22 seconds. Finally, we considered the electron pitch angle distribution at 314~eV from the Solar Probe ANalyzer-Electron \citep[SPAN-E,][]{whittlesey_solar_2020}, whose cadence is 28 seconds. 

\subsection{Coordinate systems}

Throughout our analysis, we used different coordinate systems. RTN coordinates were used for the analysis of switchbacks and their boundaries: $R$ is radial and points away from the Sun, the tangential $T$ component is the cross-product of the solar rotation vector with $R$; and the normal $N$ component completes the right-handed set and points in the same direction as the solar rotation vector. 

For the analysis of the current sheets, we expressed the magnetic field and proton velocity measurements in the local current sheet coordinate system, LMN: $L$ being the main direction of the current sheet, $M$ the out-of-plane direction, and $N$ the normal. We followed the same hybrid minimum variance method as described in \citet[][Sect. 2.3]{phan_parker_2020}, using the two points enclosing the abrupt change in $\mathrm{B_R}$ as references for the calculations.

%%%%%%%%%%%%%%%%%%%%%%%%%%%%%%%%%

\section{Evidence for magnetic reconnection at the boundaries of magnetic switchbacks}\label{sec:analysis}

Here, we present the analysis of three magnetic switchbacks showing evidence for magnetic reconnection at their boundaries. The selection of the events was made through visual inspection of a large list of switchbacks. We do not aim here to present an exhaustive detection of such events. They seem rare, and the cases we report here are the most striking ones.

\subsection{Reconnection exhausts with strong guide field}
\label{sec:event_1&2}

The two switchbacks that are presented in this section show strong similarities and are thus grouped together here. As we detail, strong pieces of evidence for magnetic reconnection are detected at both edges of these switchbacks.

\subsubsection{Event 1 on November 2, 2018 at 13:05:09~UT}

Figure~\ref{fig:context_event_1} shows the main features observed during the crossing of the switchback constituting event~1, observed on November 2, 2018. 
The leading edge crossing started at 13:05:09~UT and the trailing edge one started at 13:15:11~UT. These crossings, which last $5.6\pm 0.5$~s and $7.9\pm 0.5$~s , respectively, are highlighted with a light red shaded area in Fig.~\ref{fig:context_event_1}. These windows correspond to the abrupt changes in $\mathrm{B_R}$. For this switchback, the total crossing time was about 10 minutes.

We note that $\mathrm{B_R}$ went from about -42~nT to a maximum of 8~nT inside the structure. We note that $\mathrm{B_R}$ dropped again to a negative value inside the structure, that is -52~nT, which is approximately the value reached after the crossing of the structure. The average radial proton velocity during this time window was 320 \kms, with a clear enhancement inside the switchback with $\mathrm{v_R}$ varying from about 312 \kms \, before the switchback, up to 352 \kms \, inside. This velocity enhancement is smaller than the local Alfv\'en velocity, even though the minimal changes in $|\mathrm{B}|$ and strong correlation between $\mathrm{B_R}$ and $\mathrm{v_R}$ evoke a quasi-Alfv\'enic perturbation.

The electron pitch angle distribution (PAD) at 314~eV shows higher fluxes at a pitch angle of 180\deg \,(anti-field aligned) before, during and after the crossing of the structure. Such a constant PAD behaviour is a typical feature of a switchback. It reveals the local kink in the magnetic field structure (see Sect.~\ref{sec:intro}).

Jets, mainly in the R and N components of the proton velocity, are clearly visible at both edges of the structure. 
In the two sets of plots at the bottom of Fig.~\ref{fig:context_event_1}, we show close-ups of the two boundaries treated as current sheets. The magnetic field and proton density are now expressed in the respective local current sheet coordinate system: LMN. In this frame, we now compare the evolution of $\mathrm{B_L}$ and $\mathrm{v_L}$, that is the magnetic field and proton velocity along the current sheet, at the two boundaries of the switchback. 
We note that for both current sheets, the guide field $\mathrm{B_M}$ was strong compared to the reconnecting field $\mathrm{B_L}$: 2.3 and 1.3 times higher, respectively.

At the leading edge, the proton jet now seen in the main direction of the current sheet has a $\Delta v_L = -52$ \kms, which in an absolute value represents approximately $50~\%$ of the local Alfv\'en velocity. For this boundary, we do not notice any significant density or radial temperature increase spanning multiple data points.

At the trailing edge, the proton jet has a $\Delta v_L = + 107$ \kms, that is $113~\%$ of the local Alfv\'en velocity. For this boundary there is a significant increase in the proton density of $20~\%$. As for the other boundary, there are no significant variations in the radial temperature.

We now compare the relative variations of $\mathrm{B_L}$ and $\mathrm{v_L}$ in order to check whether there are correlated variations. This analysis is displayed in Fig.~\ref{fig:correlation_boundaries_event_1} for both the leading and trailing current sheets. Here, $\mathrm{B_L}$ was decimated to the cadence of $\mathrm{v_L}$. Next, we computed the Pearson correlation coefficient between $\mathrm{B_L}$ and $\mathrm{v_L}$ for time windows that delimit the boundaries. For each current sheet, these time windows were chosen in such a way as to maximise the correlation coefficient. 

For the leading edge, we find a positive correlation of 0.65 between $\mathrm{B_L}$ and $\mathrm{v_L}$, followed by an anti-correlation of -0.79. For the trailing edge, we observe the same succession: a positive correlation of 0.97 followed by an anti-correlation of -0.87.

A change of sign in the correlation between $\mathrm{B_L}$ and $\mathrm{v_L}$ is expected for reconnection exhausts (see Sect.~\ref{sec:intro}). Together with the associated proton jet they provide direct evidence for reconnection at the boundaries of the switchback. We note that the trailing edge current sheet had already been reported by \citet{phan_parker_2020} as a regular solar wind event (i.e. not linked to a particular structure as an HCS crossing or ICME).

\subsubsection{Event 2 on November 1, 2018 at 23:23:04~UT}

Event~2 was registered on November 1, 2018. The crossing of the leading edge occurred at 23:23:04~UT and lasted $3.8\pm0.5$~s. PSP crossed the trailing edge at 23:25:01~UT for $27.7\pm0.5$~s. This second boundary is large and has several sub-structures. The crossing of the switchback itself lasted for 1~minute and 50 seconds in total, which is considerably faster than for event~1. Event 2 is presented in Fig.~\ref{fig:context_event_2}. 

We note that $\mathrm{B_R}$ went from approximately -30~nT before the switchback, to +37~nT inside the structure. The magnitude of proton velocity also increased inside the switchback by about 10 \kms compared the velocity before the crossing. The PAD properties remain constant throughout the structure as for event~1, which is expected for magnetic switchbacks. We note that this structure is embedded in a larger switchback, which we present in Fig.~\ref{fig:context_event_2_large}. The crossing of this larger switchback started at 23:16:59~UT and ended at 23:25:29~UT. Analysing the properties of this complex structure is beyond the scope of the present paper, but we note that this structure shows overall an increase in $|\mathrm{B}|$ and decrease in the proton density with anti-correlation between these two variables. This is consistent  with the slow magnetosonic perturbations reported in \citet{larosa_switchbacks_2020}.

Going back now to the sub-structure constituting event 2, from the measurements in the RTN frame only, it is already clear that both boundaries show proton jets, with an increase and a decrease in velocity, respectively.
In the local current sheet coordinate system, it appears that these current sheets also had a strong guide field: 2.3 and 2.6 times the reconnecting field, respectively.
The jet at the leading edge has a $\Delta v_L = + 22$ \kms, that is $26~\%$ of the local Alfv\'en velocity. For the trailing edge, $\Delta v_L = - 52$~\kms, that is $60~\%$ of the local Alfv\'en velocity.
No clear density nor radial temperature variations were observed, except for the $41\% $ proton density increase at the trailing edge. We present the correlated variations of $\mathrm{B_L}$ and $\mathrm{v_L}$ in Fig.~\ref{fig:correlation_boundaries_event_2}. At the leading edge, the positive correlation (0.95) is followed by an anti-correlation (-0.80). The same occurs at the trailing edge, with a correlation (0.96), followed by an anti-correlation (-0.76).

We conclude that both boundaries of this switchback also undergo magnetic reconnection. We note that these events were also reported by \citet{phan_parker_2020} as regular solar wind events. 

We similarly analysed the leading edge of the larger structure presented in Fig.~\ref{fig:context_event_2_large}. No evidence for reconnection was found in that current sheet.

\subsection{Event 3 on November 1, 2018 at 18:18~UT}\label{sec:event_3}

First, we present the context for event 3 in Fig.~\ref{fig:context_event_3}. This magnetic structure is close to a full reversal of the magnetic field, with a rotation of $\mathbf{B}$ of approximately 165\deg. We note that $\mathrm{B_R}$ goes from about -30~nT before the switchback to +40~nT inside and then back, close to its original value. For this structure there is also a full reversal of $\mathrm{B_T}$ and a large deflection in $\mathrm{B_N}$. The inner part of the switchback is highly structured, with several smaller deflections in the field. However, we do not consider these sub-structures and focus on the structure as a whole. 

PSP crossed the leading and trailing edges of the switchback at approximately 18:21:28~UT and 18:34:39~UT, respectively. These crossings lasted for about $6.0\pm0.5$~s and $16.2\pm0.5$~s, respectively. The total crossing time of the structure is about 13 minutes. For both boundaries, there is a strong decrease in the magnitude of the magnetic field by about $90\%$, down to 3~nT.

The average radial proton velocity during this time window is 342 \kms, with a clear enhancement inside the switchback with $\mathrm{v_R}$ varying from about -40 \kms to 40 \kms, which is an increase that is close to the local Alfv\'en velocity. The proton density inside the structure decreases from about 200~\cm3 to 150~\cm3 but shows a local increase of about $15\%$ at the trailing edge. This behaviour is consistent with the observation of typical switchbacks with magnetic decreases at their boundaries \citep{farrell_magnetic_2020}. However, here, there is no clear increase in the proton temperature. We note that alongside the strong decrease in $|\mathrm{B}|$, there is no dramatic increase in the proton density and almost none in the proton temperature. Consequently, $\mathrm{\beta_{p}}$ is high at the switchback boundaries, reaching 68 and 113 at the points of reversal, respectively. 

We note a suppression of the strahl at both  boundaries. It is particularly striking at the trailing edge where around the minimum in $|\mathbf{B}|$, the flux is weak and not field-aligned. Since the magnetic field direction is changing quickly with respect to the sampling period of SPAN-E, one has to be careful with the interpretation. Meanwhile, we also see a decrease in the electron flux in the two neighbouring bins of the magnetic dip. Additionally, we meticulously checked that the rotation of the magnetic field in these bins is properly captured in the data used to compute the PADs. From these analyses, we conclude that the strahl dropouts are real and are not due to data processing.

We now focus on the boundaries of the switchbacks. As described above, they share  similarities:
large magnetic field dips; changes in the strahl; a very high value of $\beta\mathrm{_p}$; and a small increase in the proton density.
However, the configuration of the magnetic field at the trailing edge is very different as it shows a full reversal of the three components with a large magnetic shear of 168\deg. Combined with the strong decrease in magnitude, this could indicate that PSP was very close to an X-point. 

For the leading edge, the current sheet crossing lasts about 0.7~second, which is not enough to resolve a possible reconnection process (only three data points in the SWEAP measurements). However, we notice a velocity enhancement, although on only one point, which may be the signature of a reconnection jet. For this point at the middle of the current sheet, $\Delta v_L = 35 $ \kms, which is about $75 \%$ of the local Alfv\'en velocity. Although not resolved, the presence of this jet is consistent with a change in sign for the correlation between $\mathrm{B_L}$ and $\mathrm{v_L}$, which is expected for reconnection exhausts. We finally note that this current sheet is located in a relatively modest magnetic dip (decrease in $|\mathbf{B}|$ by $22\%$) preceding the large dip at the leading edge of the switchback. The guide field was strong: 1.4 times that of the reconnecting field.

The analysis of the compared evolution of $\mathbf{B}$ and $\mathbf{v}$ for the trailing edge is shown in Fig.~\ref{fig:correlation_boundaries_event_3}. In the first time window, ending approximately at the minimum of $|\mathbf{B}|$, there is a clear correlation of $\mathrm{B_L}$ and $\mathrm{v_L}$ (correlation coefficient of 0.9). The second time window goes from the minimum of $|\mathbf{B}|$ to the end of the boundary. In that case too, there is a change in sign of the correlation coefficient (-0.44), which indicates an anti-correlation. This weak anti-correlation can be explained by the noisy evolution of $\mathrm{v_L}$. It is also symptomatic of the fact that there is no proper velocity enhancement in order to call it a jet. We note here that we also did the analysis with the regular MVA method \citep{Sonnerup1998} which can be more suited for large magnetic shear, but this did not improve the correlation values. To conclude, although these correlated changes of $\mathrm{B_L}$ and $\mathrm{v_L}$ are consistent with the observation of the crossing of a reconnection exhaust, no proton jet is detected near the inversion of $\mathrm{B_L}$.

We note that the ratio of the guide field $\mathrm{B_M}$ to the reconnecting field $\mathrm{B_L}$ is weak with a value of $0.05$. This is quite different from events~1 and~2, and the leading edge of event~3, for which the guide field was strong and, consequently, $|\mathbf{B}|$ remained almost unchanged during the current sheet crossings.

For event 3, whistler wave signatures were detected for both boundaries between 10-80~Hz and 20-100~Hz, respectively. These frequencies fall between $\mathrm{f_{lh}}$ and $\mathrm{f_{ce}}$, the lower hybrid and electron cyclotron frequency. The whistler waves detected around the trailing edge are shown in Fig.~\ref{fig:correlation_boundaries_event_3}. The BPF data reveal that these waves were encountered at the reconnection mid-plane, between 6 and 16 seconds after the minimum of $|\mathbf{B}|$. 

By applying the singular value decomposition technique \citep{santolik_singular_2003} to the cross-spectral data, we found that these waves have a high planarity (0.8) and ellipticity (0.9). Considering $\mathbf{B}$ at the times of the detected whistler wave activity in the BPF, that is after the magnetic dip, we computed the angle between the $k$-vector and  $\mathbf{B}$ and found $\mathrm{\theta} = 6\deg \pm 180\deg$. These waves are therefore quasi-parallel. Quasi-parallel whistlers are often observed in the vicinity of reconnection regions \citep[e.g.][]{wei_cluster_2007,fujimoto_whistler_2008, graham_whistler_2016,huang_occurrence_2017, voros_mms_2019}. No such clear wave packet was observed at the leading edge.

In conclusion, for the trailing edge of event~3, the correlated changes in $\mathbf{B}$ and $\mathbf{v}$ are consistent with magnetic reconnection. Moreover, the clear strahl dropout in the area encompassing the current sheet may indicate a disconnection from the Sun. The fact that PSP encounters quasi-parallel whistler waves at the boundary of the current sheet, right after the steepest variation in $\mathrm{B_L}$, could indicate that the spacecraft was close to the ion-diffusion region on that side of the current sheet \citep[e.g.][]{voros_mms_2019}. We checked the trajectory of the probe in the LMN frame and found that its 2D projection in the LN plan is oblique. This can support the assumption that the probe got closer to the X-line while crossing the current sheet, as it was already suspected from the inversion of three components of the magnetic field and the large magnetic dip.
The promixity of the X-line could explain why no reconnection jet was observed.

On the other hand, in computing the distance to the X-line, we found that the probe should have been quite far from it.
Using the proton velocity along the normal direction to the current sheet ($<\mathrm{v_N}> = 190$ \kms) and the crossing time ($16.2\pm0.5$), we evaluated the thickness of the current sheet to be $\delta L = 3083$~km. This translates into 186 ion-inertial lengths. Using a standard 0.1 reconnection rate \citep[e.g.][]{birn_geospace_2001, liu_why_2017}, we found $\frac{\delta L/2}{0.1} = 15416$~km, that is 945 ion-inertial lengths. Such a large number contradicts the previous arguments suggesting that PSP was moving close to the reconnection site. In addition, we found no Hall perturbations in the out-of-plane component of the magnetic field. Close to the X-line we should witness a bipolar variation in $\mathrm{B_M}$ around the reversal in $\mathrm{B_L}$ \citep{oieroset_situ_2001}, which is not the case here.

However, looking at the evolution of $\mathrm{B_L}$ (see Fig.~\ref{fig:context_event_3}), it is clear that the steep decrease in the magnetic field only happens at the very end of the crossing. If the probe would have gotten closer to the X-line by the end of the crossing, it would not have been possible to resolve any Hall perturbations.

\begin{table*}
	\caption{Summary of the individual indications for magnetic reconnection.}
	\label{table:summary_reconnection}
	\centering
	\begin{tabular}{c c c c c c c}
		\hline\hline
		\multirow{2}{*}{Event} & {Leading/Trailing} & Correlated/Anti-correlated & \multirow{2}{*}{Jet} & Strahl & \multirow{2}{*}{Whistlers} \\
		 & edge & $B_L$ \& $v_L$ Changes &  & suppression & \\
		\hline
		 1 & Leading & yes  & yes & no & no \\
		 1 & Trailing & yes & yes & no & no \\
		 2 & Leading  & yes  & yes & no & no \\
		 2 & Trailing & yes & yes & no & no\\
		 3 & Leading & yes (not resolved)  & yes (not resolved) & likely & no  \\
		 3 & Trailing & yes (weak anti-correlation)  & no & yes & yes  \\
		\hline

	\end{tabular}
\end{table*}

\begin{table*}
	\caption{Characteristics of the reconnection events detected in the current sheets (CS) constituting the switchback boundaries.}
	\label{table:summary_events}
	\begin{adjustbox}{width=\textwidth}
	    \begin{tabular}{c c c c c c c c c c c}
		\hline\hline
		\multirow{2}{*}{Event} & {Leading/Trailing} & |$\mathbf{B}$| dip & \multirow{2}{*}{Guide field} & Rotation $\mathbf{B}$  &  \multirow{2}{*}{$\beta\mathrm{_p}$} & $j$ & Density & CS Thickness & Distance to the X-line & Jet velocity \\
		 & edge &  [\%] &  & [\deg]  &  & [$\mathrm{nA/m}^{2}$] & increase [\%] &  [$\mathrm{d_i}$] ([km]) & [$\mathrm{d_i}$] ([km]) & [$\% \mathrm{v_A}$]\\
		\hline
		 1 & Leading & 6  & 2.3 &  58 & 0.4 & 47.7 & -  & 60 (1004) & 301 (5019) & 50 \\
		 1 & Trailing & 27  & 1.3 & 81 & 0.7 & 36.9 & 25 & 107 (1770) & 539 (8851) & 113 \\
		 2 & Leading  & 16  & 2.3 & 67  & 0.9 & 75.5 & - & 40 (640) & 203 (3205) & 26 \\
		 2 & Trailing & 12 & 2.6 &  60 &  0.9 & 18.3 & 37 & 164 (2569) & 821 (12847) & 60 \\
		 3 & Leading & 22 & 1.4 & 61 & 2.3 & 181 & - & 10 (155) & 50 (776) & 75 \\
		 3 & Trailing & 90  & 0.05 & 168 & 112.6 & 15.2 & 15 & 186 (3083) & 945 (15416) & - \\
		\hline

	    \end{tabular}
	\end{adjustbox}
\end{table*}

%%%%%%%%%%%%%%%%%%%%%%%%%%%%%%%%%%%%%%%%%%%%%%%%%%%%

\section{Summary and discussion}\label{sec:conclusion}
 
In this paper, we report the observation of magnetic reconnection at the boundaries of three magnetic switchbacks observed by PSP. These events were detected during the first encounter of PSP in November 2018. The signatures associated with magnetic reconnection at the switchback boundaries are summarised in Table~\ref{table:summary_reconnection}. The main characteristics of these current sheets are listed in Table~\ref{table:summary_events}, which namely include the following: magnetic dips, the guide field, magnetic shears, an increase in $\beta\mathrm{_p}$, an increase in the current density and plasma density,  the current sheet thickness, distance the to the X-line, and the reconnection jet velocity.
 
Most of the events that we identified are reconnection exhausts with a significant guide field and a moderate magnetic shear. However, we also identified one event with possible indications of a quasi anti-parallel reconnection (magnetic shear of 168\deg) with a strong dip in the magnitude of the magnetic field. This suggests that a wide range of reconnection properties may occur at switchback boundaries.

Events~1 and~2 are switchbacks with high guide field magnetic reconnection at both their boundaries. These four current sheets show a typical signature of reconnection exhaust with the following: a correlation and then anti-correlation of the magnetic field and proton velocity in the main direction of the current sheet, and clear proton jets with velocities ranging from $26\%$ to $113\%$ of the respective local Alfv\'en velocity. 
It is not uncommon for reconnection jets to be only at a fraction of the Alfv\'en velocity. There are several explanations for this as detailed in \citep{phan_parker_2020}.
These events have a strong guide field as compared to the reconnecting field. Conversely, they have modest dips in the magnitude of the magnetic field \citep[also noted in][and references therein]{phan_parker_2020}. Finally the angle made by the magnetic field at the two sides of the current sheet is moderate, that is 60\deg to 80\deg. Similar characteristics are found at the trailing of event~3, which shows a good indication of reconnection, even though the reconnection jet was not properly resolved.

Event~3 is a switchback with large magnetic field dips of about $90\%$ at both its leading and trailing edges. The trailing edge current sheet shows several pieces of evidence for anti-parallel reconnection. Correlated and anti-correlated variations in the magnetic field and proton velocity upon entry and exit were found, as for the other events. Moreover, the electron strahl was suppressed in this area, indicating a possible disconnection from the Sun. However, the absence of a reconnection jet prevents us from fully concluding on this matter. While the presence of quasi-parallel whistler waves at the boundary of the current sheet, right after the steepest variation in $\mathrm{B_L}$ (and thus the highest variation in the current) could point to a close approach to the X-line, the distance from the X-line that we computed reveals that the probe was far enough from the X-line to be able to detect such a jet at some point of the crossing.
Since reconnection is a 3D process, with only one spacecraft crossing the structure, we may not have totally uncovered its geometrical complexity.

One striking feature of this last event was the large magnetic dips at both boundaries.
\citet{farrell_magnetic_2020} show that switchbacks with magnetic field dips at their boundaries are quite common. As conjectured by these authors, such dips may be created by the diamagnetic currents that are due to the magnetic shear and gradient in density at the boundaries. This has also been noted by \citet{krasnoselskikh_localized_2020}. 
We also evaluate the total current density of $j\simeq \delta B / (\mu _{0}\delta L)$ and find comparable values to those reported by \citet{krasnoselskikh_localized_2020} (see Table~\ref{table:summary_reconnection}.)
In the events that we report, the rotation of the magnetic field ranges from 58\deg to 168\deg \, and the current density ranges from 15.2 to 181~$\mathrm{nA/m}^{2}$.
It is quite natural that the current density for larger deflections of the magnetic field are stronger. Thus, one can suggest that this type of reconnection mainly occurs across the boundaries of strong magnetic field deflections.

The reconnection events we report were the most striking ones we could find during PSP's first encounter. So far, they seem to be rare but a more systematic search is needed to quantify their occurrence in the solar wind.
On the other hand, magnetic switchbacks seem to occur more frequently in the pristine solar wind. Why they seem more scattered closer to 1 AU has yet to be determined.

The reconnection events we found are all at a distance close to 50 solar radii. We also notice that the velocity increase inside the switchbacks that we analysed is moderate compared to the strong jets that were observed close to the perihelion. On the one hand, we could conjecture that the reconnection at the boundaries destabilise the structure and causes the velocity enhancements to decay.
On the other hand, the statistical study of switchbacks near the perihelion presented in \citet{larosa_switchbacks_2020} show that the magnitude of these velocity enhancements is only a few percent of the velocity outside the switchbacks. The velocity enhancements of the structures presented here are consistent with their distribution.
However, one could wonder whether strong velocity enhancements would prevent the reconnection for happening at the boundaries of switchbacks. Via simulations, \citet{swisdak_diamagnetic_2003} demonstrated that diamagnetic drifts suppress reconnection when the velocity of the moving structure becomes comparable with the Alfv\'en velocity. They formulated that the reconnection rate would depend on the magnetic shear and on the $\beta$. Such parameters would be interesting to explore in future studies.

Overall, one can conjecture that the reconnection process happening at switchback boundaries may contribute to the blending of the structures with the regular wind, through a reconfiguration of the magnetic field, which would eventually dissipate the structures. Magnetic reconnection could erode the switchback boundaries as they travel, mostly affecting strong deflections. Smaller deflections would be less likely to undergo such processes. However, small switchbacks would probably disappear at large distances from the Sun anyway due to the interpenetration of the plasma outside and inside the structures and the restoring force of the magnetic field.

Many more questions arise from these observations. In particular, magnetic reconnection at switchback boundaries could contribute to the local heating of solar wind plasma. However, more investigations, in particular statistical studies, would be needed to evaluate their contribution to the global energetics of the solar wind.

\begin{acknowledgements}
The authors thank the anonymous referee for their constructive comments.
C.F., V.K., T.D., M.K and A.L. acknowledge funding from the CNES. O.A. and V.K. were supported by NASA grant 80NSSC20K0697. O.A. was partially supported by NASA grants 80NNSC19K0848, 80NSSC20K0218, and NSF grant NSF 1914670. S. D. B. acknowledges  the support of the Leverhulme Trust Visiting Professorship program. Parker Solar Probe was designed, built, and is now operated by the
Johns Hopkins Applied Physics Laboratory as part of NASA’s
Living with a Star (LWS) program (contract NNN06AA01C).
Support from the LWS management and technical team has
played a critical role in the success of the Parker Solar Probe
mission. All the data used in this work are available on the
public data archive NASA CDAWeb (https://cdaweb.gsfc.nasa.gov/index.html/).
Figures were produced using Matplotlib \citep{Hunter:2007}.
\end{acknowledgements}

\bibliographystyle{aa}
\bibliography{paper_reconnection_SB}

\end{document}